*Article*

# How to Engage Active Pedagogy with Physics Faculty: Watch Out for Powerlessness

Andria C. Schwortz [1,2,*], Michael Frey [1] and Andrea C. Burrows Borowczak [3]

[1] Department of Natural Sciences, Quinsigamond Community College, Worcester, MA 01606, USA
[2] Physics Department, College of the Holy Cross, Worcester, MA 01610, USA
[3] School of Teacher Education, University of Central Florida, Orlando, FL 32816, USA; andrea.borowczak@ucf.edu
* Correspondence: aschwortz@holycross.edu

**Abstract**

Despite the large body of research showing that students in STEM classes at all levels learn better via active learning than they do via lecture, post-secondary physics and astronomy (P&A) faculty members continue to primarily use teacher-focused, lecture pedagogy in their classes. Methods include answers from eight faculty members, and interviews with five faculty members who self-identified as primarily using lecture were conducted to determine their perceptions of why they use lecture. During analysis coding, results show that an unanticipated theme not sufficiently represented in the pre-existing literature rose to the forefront: that many of these faculty members feel the decision of pedagogy is out of their control. In conclusion, a grounded theory was developed and is proposed herein that these faculty feel a sense of powerlessness. Reasons offered include administrators often make decisions based on the financial needs of the school, which then force the faculty into using lecture as their primary pedagogy. Implications include that providing professional development in active pedagogies may not be sufficient to help faculty members change pedagogy, as they may need to be convinced that they have the power to make change and use student-centered, active learning pedagogies within their own individual constraints and settings. Understanding that some instructors may feel powerless in choosing how to teach is an important step for professional development providers toward ensuring that faculty have a voice and can choose the best teaching methods for their classrooms.

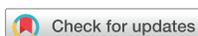

## 1. Introduction

Student-centered pedagogy (or active learning) has been shown to be a more effective pedagogy than teacher-centered pedagogy (or lecture) in all fields and levels of education (Jarilkapovich, 2025), including physics and astronomy (Schwortz & Burrows, 2021; Burrows et al., 2016; Hake, 1998). The incorporation of modern technology into education also opens more opportunities for new forms of student-centered pedagogy within science classes of all levels (Jarilkapovich, 2025; Schwortz et al., 2023). Recent evidence shows that pre-K-12 science education is increasingly moving towards student-centered pedagogy (e.g., Leung, 2023). In addition, many nations' standards are encouraging pre-K-12 teachers to move towards more student-centered pedagogy such as authentic science experiences (Schwortz & Burrows, 2021).

Despite this, at the post-secondary level many physics and astronomy (P&A) faculty continue to use lecture and seem to resist reform efforts (Bland et al., 2007; Bligh, 2000; The Boyer Commission on Educating Undergraduates in the Research University, 1998; Mitchell, 2009; Waldrop, 2015; Zakirman et al., 2019). This is consistent whether they are teaching to STEM majors (The Boyer Commission on Educating Undergraduates in the Research University, 1998), education majors (Burrows et al., 2016), or non-majors in general education classes (Mitchell, 2009). While extensive research exists on how to assist educators at all levels in changing from teacher-centered to learner-centered pedagogy (e.g., Sansom et al., 2023), there is limited evidence as to why many college faculty members continue using teacher-centered pedagogy, and any potential reasons (e.g., class size, space limitations, lack of training, the ease of using lecture, tradition) remain speculative, without supportive evidence.

In trying to investigate whether these and other speculated causes are actually evidenced by the physics faculty members themselves (to be discussed in a future paper), the authors discovered an emergent theme that was not anticipated by the existing literature. In this study, a number of the participants indicated that what looked like a choice of pedagogy to outsiders was actually experienced as being dictated by factors outside of their control. As a result, this paper provides a grounded theory of how eight P&A faculty were engaged in a study and expressed their feelings of powerlessness (in varying degrees) over the pedagogy they use in their courses. In short, the authors showcase data and answer questions about how these faculty describe their sense of powerlessness over their pedagogy.

*1.1. Literature Review*

While many researchers speculate as to why post-secondary faculty choose to use lecture, few studies have actually focused on finding evidence of these reasons. For example, Sansom et al. (2023) examined why STEM post-secondary faculty members switch from instructor-centered to student-centered pedagogies, as well as reasons why they find doing so difficult, but they acknowledged a limitation of the study was that participants had already self-selected as wanting to change their pedagogy (see Section 4.1). As a result, sources are often primarily speculating about the factors they suggest, without actual evidence. These seven speculated causes were used as a priori themes to begin this study, including (1) instructor's time and effort, (2) lack of training, (3) tradition and survivorship bias, (4) student learning considerations, (5) assessment of instructor, (6) class size and space limitations, and (7) additional administrative considerations. These causes are discussed in the following paragraphs.

Instructor-centered teaching or lecture is perceived to be less time-consuming for faculty to prepare to teach, and it takes less time during class to deliver the content (Burrows et al., 2016; Malicky et al., 2007; Waldrop, 2015; Zakirman et al., 2019), in addition to allowing for a more "coherent and logically organized presentation" (Malicky et al., 2007, p. 326). The concern about class time can compound when there is a need to remediate pre-collegiate knowledge and skills (The Boyer Commission on Educating Undergraduates in the Research University, 1998), as is often the case in introductory physics classes. And the preparation time for class is competing with the time that faculty have available to perform research (Mitchell, 2009; Waldrop, 2015).

The time required for this preparation can be further increased when faculty have received a minimal amount of training in alternative pedagogies (Zakirman et al., 2019). This is then reinforced by a survivorship bias, with faculty thinking that, since they thrived as a student in a lecture setting, their students should as well (Bland et al., 2007). In addition, if faculty do not know how to implement active learning well or do not believe in using it, they can actually deter student learning (Breslow, 2010; Waldrop, 2015). Between this risk and the many students who (incorrectly) believe they learn more via lecture (Flaherty, 2023; Kapon et al., 2010), faculty may feel it is safer to use lecture. There is also the risk that faculty members using student-centered pedagogies receive worse evaluations both from supervisors and students (Burrows et al., 2016; The Boyer Commission on Educating Undergraduates in the Research University, 1998; Flaherty, 2023; Kapon et al., 2010; Tadesse et al., 2021).

Additionally, both large class sizes (number of students) and physical classroom spaces (the area of the room, and the layout of the seats or desks) are cited as factors that lead faculty members to teach via lecture (Burrows et al., 2016; Mitchell, 2009; Tadesse et al., 2021; Zakirman et al., 2019). While class size and physical space are two separate factors, if a small number of students are placed in a large room, it could differ from a large number of students placed in a small room. These factors may or may not be independent.

There are other factors that can influence pedagogy. For example, many administrators are concerned about finances and focus on cost-efficient delivery of classes. This can influence the adoption of online education (Kortemeyer et al., 2023), larger class sizes (The Boyer Commission on Educating Undergraduates in the Research University, 1998), and less of a focus on research (Waldrop, 2015). When administrators receive mixed messages about pedagogy, such as students themselves believing (most likely incorrectly) that they learn more via lecture than active learning (Flaherty, 2023), administrators may feel that making decisions based on finances is a reasonable option.

However, many of these concerns can be ameliorated via known teaching tools and pedagogies. For example, professionally published workbooks containing lecture tutorials and conceptual tasks (e.g., TIPERS, Tasks Inspired by Physics Education Research, Hieggelke et al., 2013; Astronomy Activity and Laboratory Manual, Hirshfeld, 2018) allow instructors to add student-centered activities to even large lecture classes with minimal preparatory time. For faculty with low exposure to student-centered pedagogies, over a thousand colleges and universities have centers of teaching and learning that teach professional development (PD) workshops for faculty including student-centered pedagogies (Wright, 2019). In the fields of P&A specifically, many professional conferences include workshops on student-centered pedagogy, and some are even being presented at multiple conferences (e.g., Baylor et al., 2025, 2026).

When reviewing the literature, the authors found that articles focused on various factors influencing faculty provided speculation without evidence. The researchers of these studies also did not investigate the internal mental mechanism by which these factors influenced the faculty. As a result, the theme of powerlessness discussed in this research study is not sufficiently addressed in the existing literature about STEM faculty pedagogies nor in the existing literature about STEM instructor professional development with student-centered pedagogies.

Although much of the literature discusses power and agency in the classroom, especially of students (e.g., Jones, 2015), most studies assume that faculty have the power to make "pedagogic decisions... influenced by many factors" (Burridge, 2018). However, few studies focus on the faculty's lack of agency to choose student-centered pedagogies within a STEM context, let alone P&A specifically. A scan of the existing literature on powerlessness in choosing STEM student-centered pedagogies was performed via Google Scholar (https://scholar.google.com/) using search terms such as "powerlessness pedagogy STEM" and "powerlessness in STEM teachers." Within these search terms, research into pedagogical power regarding active learning generally investigates the sense of student powerlessness in teacher-centered classrooms (e.g., Burrows et al., 2016; Chase, 2020; Freeworth, 2021; Ho, 2024).

The education articles that investigate the faculty member's sense of powerlessness generally do not focus on the applicability in pedagogy choice. Other aspects of powerlessness that are investigated include the COVID crisis (Thacker et al., 2022), the context of urban education (Goss, 1975; Vontress, 1963), and teachers' concern that they are not ready to teach modern technology (Mlangeni & Seyama-Mokhaneli, 2024). Only one article investigated a faculty member's sense of powerlessness over their pedagogy, but the purpose was in fact opposite of what was primarily discovered in this article: the faculty member preferred a student-centered approach but was forced to use an instructor-centered curriculum (Gordon, 1996).

*1.2. Gap in the Literature*

While there is a broad need to focus STEM education research on the post-secondary level (Burrows et al., 2016; Tadesse et al., 2021), there is also the specific need to understand why faculty continue using only traditional pedagogies and how these views about pedagogy can be modified (Duit et al., 2014; Mitchell, 2009). Unfortunately, research into the choice of pedagogy by post-secondary STEM faculty tends to focus on those individuals who want to make changes, while those who do not want to make changes have not been studied to the same extent (Sansom et al., 2023). The Handbook of Research on Science Education (Crawford, 2014) identifies a gap in the literature where more work needs to be completed to pinpoint "roadblocks" (p. 537) and factors that the "science teachers themselves identify as being problematic" (p. 536).

*1.3. Research Questions*

The original question motivating the project was: "Among post-secondary physics and astronomy (P&A) faculty members who teach primarily using lecture, what are the factors affecting their change to a more student-centered pedagogy?" As the project morphed into a grounded theory (see Section 2), the motivating research questions similarly changed to the following:

> "Among post-secondary P&A faculty members who teach primarily using lecture, and feel they do not have control over their pedagogy, . . .

1. How do they describe their sense of powerlessness over pedagogy?
2. To what factors or agents do they attribute the choice of pedagogy?
3. How do they perceive the process that causes them to primarily use lecture?".

*1.4. Theoretical Framework*

The researchers utilized grounded theory and constructionism as the theoretical framework for this study. They describe the phenomenon of some P&A faculty members as they experienced a sense of powerlessness over the pedagogy used in their classrooms. As such, the faculty participants' own words describing their experiences and the reality they constructed around them are presented to highlight the processes that they have lived (Koro-Ljungberg et al., 2009; Vygotsky, 1978).

Grounded theory was chosen due to the emergent nature of the interviewed faculty member's sense of powerlessness. Not only is grounded theory an effective choice when allowing a new understanding of a process to emerge (Burrows, 2011; Creswell, 2013; Corbin & Strauss, 1990), but it is also ideal for cases where the researcher is "already immersed" in the profession (Burrows, 2011, p. 37; Glaser & Strauss, 1967). Adding to the context, author 1 is a P&A faculty member who has a lived experience that complements this research study's grounded theory work.

## 2. Methods

Faculty participants were solicited via semi-public email lists and social media, focusing on post-secondary physics and astronomy (P&A) faculty members primarily (but not exclusively) located in the USA. Interviews of 30 to 60 min were conducted in Fall 2024 with participants over Zoom by authors Schwortz and Frey. The interviewers used a list of questions as a starting point and asked follow-up questions as needed to dig deeper into the ideas the participants raised. The interviews were audio and video recorded, and Zoom provided automated transcripts. These transcripts were examined after the conclusion of all interviews and corrected as needed based on the recordings.

The interview questions were designed to elicit anticipated theme. For example, participants were asked about the physical teaching space (e.g., class size, lecture hall size, aspects of "arm desks," or tables with chairs), teaching style, pedagogy training, other commitments, and more. Participants were also asked to speculate on which teaching methods their students enjoyed the most compared to those from which they learned the most effectively. Below are some example questions, with parentheticals indicating items that the interviewer noted and asked follow-up questions to elicit as needed. The full interview protocol is included in Appendix A.

1. Could you tell me a bit about your current faculty position?
2. In what modalities or settings of class do you teach? (e.g., lecture in a big hall with hundreds of students, small room with 10 students, lab, discussion section, online asynchronous [no meeting times, generally using either pre-recorded videos or written materials], online synchronous [specific meeting times, generally using Zoom or similar]).
3. How would you describe your teaching style? (Look for both a name of a pedagogy, and description of what they're actually doing, e.g., . . . )
4. (You said you use [pedagogy], can you describe what that term means to you?)
5. (What are the students doing during class, and what are you doing?)
6. How are you assessed on your teaching?
7. What teaching style or styles were used in your own education?
8. What types of training in how to teach/pedagogy instruction have you received?

There were a total of 56 questions, with approximately half intended as optional follow-up questions should the initial response not include sufficient detail. See Appendix A for the full interview protocol.

The solicitation for participants was sent to three email lists targeting P&A instructors. Fifteen individuals expressed interest in participating, and eight faculty members completed interviews. These eight interviews comprised a rich qualitative dataset consisting of a total of seven hours and fifteen minutes of recording, 446 pages of transcripts, and 107,902 words in the transcripts. Table 1 specifies the length of each individual transcript, the total, and the average.

Only one participant was female, which is not surprising due to the combination of P&A faculty members still being majority male and the small number of participants. Participants' backgrounds are not reported to protect their privacy. The researchers used a name generator to assist in selecting pseudonyms. Quotes presented herein may have been lightly edited for clarity, to conform to standard American English, or to further obfuscate potential cultural cues which could compromise participants' anonymity.

**Table 1.** List of all participants, the length of their interviews and transcripts, the average, and the total.

| Pseudonym | Interview Length (h:mm:ss) | Transcript Length (pages) | Transcript Length (words) |
|---|---|---|---|
| Andromeda | 1:01:06 | 62 | 14,891 |
| Bennett | 0:56:31 | 53 | 12,289 |
| Cody | 0:57:40 | 55 | 14,243 |
| Kevin | 1:00:07 | 70 | 17,311 |
| Lonnie | 0:24:26 | 35 | 6655 |
| Sam | 0:48:22 | 60 | 12,635 |
| Timothy | 1:06:37 | 71 | 16,150 |
| Wesley | 1:01:03 | 60 | 13,728 |
| Average | 0:54:29 | 58 | 13,488 |
| Total | 7:15:52 | 466 | 107,902 |

The transcripts were iteratively coded for themes, originally looking for the a priori themes. In addition to coding for the original a priori themes, the researchers followed both a "winnowing" (Creswell, 2013, p. 184) and "unmotivated looking" (Paulus & Lester, 2013, p. 5) process, by which the researchers would go back and forth between different transcripts to see if ideas mentioned by one participant were echoed by others. This process was engaged with an open mindset, so that not only the a priori codes were identified, but also additional potential codes were examined to see if there was a common thread throughout one interview, or between multiple interviews. These were all part of the open coding process. After initial coding, authors Schwortz and Frey discussed and crosschecked each other's coding to achieve consensus. For example, on participant "Lonnie," Frey performed the initial coding, Schwortz confirmed these codes and found additional ones, and Frey went back and confirmed the additional codes.

Throughout the stages of coding, memoing took place in multiple forms and in files shared between Schwortz & Frey. These included highlighting and adding comments to shared transcript files, notes in narrative form in a shared document, and excerpts pasted into a shared spreadsheet with dropdown checkboxes for codes, emails, and text messages. Additional verbal memoing took place via discussions between the authors over Zoom—for example, debating the meaning of an excerpt, or brainstorming about which excerpts might be related. Preliminary work was presented at regional conferences, allowing for more verbal memoing between the authors and conference attendees.

Using this approach, the researchers eventually examined 23 potential codes, including the seven a priori ones from the literature previously presented. Codes (including all a priori codes) that were used multiple times and by multiple participants were retained. Codes that did not appear in multiple excerpts were rejected, as were codes that did not provide any insight into either the original research question or the new ones developed through the grounded theory. The connections between these codes were examined and organized in a process of axial coding.

Throughout this process, one main theme kept returning and echoing throughout multiple participants: a subset of the participants felt that the "choice" of pedagogy was not actually voluntary. As discussed in Section 1.1, although power is a common theme in poststructuralist theories that critique societal structure, it has not been extensively discussed in the context of STEM instructors' decision-making processes for instructor-centered pedagogies. Therefore, the theme of powerlessness in this article emerged from the data itself via inductive reasoning.

Once this theme was identified, the authors/researchers reviewed the transcripts again, creating further memos, and making connections between the theme of powerlessness found in one interview then moving back to another transcript to crosscheck the ideas found in the previous interview (axial coding; Burrows, 2011; Corbin & Strauss, 1990). One participant also discussed the theme in an emailed communication with Author 1, who then asked the participant if he would be willing to add the text of the email to his transcript.

A process of selective coding was then followed. The assorted codes and themes were further organized and checked with the excerpts to confirm that the relationships discovered held up to examination. Among some of the faculty participants ($n = 8$), this theme of powerlessness was further developed into a theory of what P&A faculty view as the source of their powerlessness. This was first developed as a graphical version that was modified as additional excerpts clarified the authors' understanding. The second version was further checked against the excerpts and further modified. The third version had no disagreements with the excerpts and was deemed to encompass all the aspects discussed by the participants. This third version of the powerlessness theory was also developed as a text version.

Five participants strongly expressed the theme of powerlessness over pedagogy, and they are further described in Table 2, which also summarizes the number of powerlessness excerpts and number of words recorded in the excerpt transcripts.

**Table 2.** Description of the five individuals who strongly introduced powerlessness in responses. The two rightmost columns indicate the number of excerpts (contiguous quotes) containing the powerlessness theme and the approximate number of words from these excerpts (including clutter words such as "um," "well," "so," etc.). "CC" indicates community college; "FT" is full-time.

| Pseudonym | School Type | Position Type | Sex/Gender | # of Excerpts | # of Words |
|---|---|---|---|---|---|
| Bennett | Multiple | Adjunct | M | 3 | 433 |
| Kevin | CC | Adjunct | M | 2 | 192 |
| Lonnie | CC | FT Faculty | M | 2 | 164 |
| Sam | Private Catholic | FT Faculty | M | 1 | 124 |
| Timothy | Public Research | FT Faculty | M | 9 | 1206 |

Following the methodology of grounded theory, the authors then conducted additional theoretical sampling among the full eight faculty participants. The text version of the powerlessness theory was sent to all participants, as well as a follow-up question as to whether they agreed with what was written, in order to assess its accuracy and to triangulate findings.

## 3. Findings

All seven of the original a priori themes appeared multiple times across multiple participants and will be further discussed in a future paper. Some of the additional codes that appeared in multiple excerpts included "powerlessness," "demos," "interactive lecture," "already have non-lecture in lab/discussion," "foundation needed first," "too little time, too much content," "technology," and "peer/family support." Codes that were considered and rejected due to not appearing in multiple excerpts included "judging students," "teaching-focused schools vs. research-focused schools," "student engagement/focus vs. student entertainment," and "student ownership of the class." One code that was considered but rejected as not providing any insight into either the original research question or the new ones developed through the grounded theory was "hating grading"—not only did this code not repeat, but the participant who mentioned it did not connect it to either the theme of powerlessness or to any other themes.

The five faculty members (62.5% of the interviewed participants) described in Table 2 brought up feelings of powerlessness over the pedagogy used in their classroom even though they were not explicitly asked about it. Three participants expressed powerlessness as single isolated excerpts, and two others expressed this idea multiple times and more complexly. When the researchers returned to ask the participants about powerlessness explicitly via email, seven (87.5%) agreed with aspects of the concern about powerlessness. Including both the interviews and emails, all eight participants (100%) expressed powerlessness over pedagogy at least once. These are further described in the following sections.

*3.1. Powerlessness as a Theme in Isolated Excerpts*

Three faculty members (Lonnie, Sam, and Kevin) had single, isolated excerpts discussing a lack of power. Lonnie expressed that he feels empathy with the students taking the course who feel powerless. He understood their predicament from his experience as an undergraduate engineering major taking physics. Lonnie said, "I always wondered, why do they make us take physics when the engineering core has all the stuff again?". The students are in courses they are forced to take, and they are powerless to choose courses they might want to take.

Sam experienced his helplessness in the context of his own instruction. He said that he usually teaches via lecture, but, in one instance, he instead used a flipped classroom modality for a single class. "Flipped classroom" is a student-centered pedagogy popular in physics where students are introduced to the content via reading and videos outside class time, and then class time is spent on problem-solving, labs, or other active learning. He explained that using the flipped modality was not a voluntary choice on his part but that it occurred when "one of my faculty departed, and I kind of needed to keep some of the classes he was teaching going. . . . [The previous professor] had it set up as flipped classroom, so I kept that." Sam expressed that, for him, lecture is the default, and anything else is something that he is forced into by the specific population of students and the precedent from prior faculty members. For Sam, the powerlessness occurred when he was forced into a student-centered pedagogy.

On the other hand, Kevin had an interest in using more student-centered learning in his astronomy classes but pointed out that his college's lack of resources was a challenge. When asked whether he used telescopes with his astronomy classes, Kevin replied that his college "has no telescopes or facilities, or anything", so he encouraged them to attend "several facilities nearby" that offer "public telescope viewing nights."

Among these three participants, the author researchers found no broad consensus as to the role of powerlessness in their teaching, yet they are presented for data transparency and a further data set. Powerlessness does play a role in Lonnie's view of his students, and Sam and Kevin's choice of pedagogy; however, the next two participants expressed a larger and more coherent role of powerlessness.

*3.2. Powerlessness as a Theory*

Two other faculty, Bennett and Timothy, showed signs of having a broad model of the interconnectedness of different aspects of academia that forced faculty into the situation of using lecture, potentially against their will. They both felt that financial needs drove the choices of administrators, who then made choices of classrooms and class sizes, which inevitably forced faculty members to teach via lecture. Bennett saw this financial driver as being the corporate culture of the wider society, which put pressure on families and then influenced the schools.

> *The corporate culture is heavily influencing... the education system... Everywhere huge money, huge fee structure and making them professionally powerful in search of good*

> *lucrative jobs, campus placement. The moment they take admission in a prominent institution the pressure starts building upon them. Parents are under pressure. Students are under pressure from the corporate houses. Only money culture is there. In a school, college or university, everything is under corporate pressure which derails the entire system. So this corporate culture is killing education.*

Regarding Timothy, the authors first took note of his discussion of class size, as this was one of the anticipated factors in the decision of pedagogy. For example, Timothy said, "I've taught astronomy lectures... with over 300 students... I don't think they let [physics classes] go above 150, but you know it's still a hell of a lot." As the researchers went back and forth between his excerpts, we noticed that he brought class size up multiple times, each time emphasizing that someone other than himself assigned him the class size, and attributing this to a nebulous "they": "I've had well, I think it was astronomy lecture one time with just 30 students. They still put me in the big lecture hall that seats 300, and that creates an issue."

Combing back through these quotes and others, the authors saw that Timothy attributed the class size as a causative factor in his pedagogy choice. As he said above, teaching in a lecture hall with "just 30 students... creates an issue." Timothy further expanded upon this in the following excerpt. "I think it's partly because of the classroom, you know. If you're in a big lecture hall, group solving is a little more difficult than when you have the students at the tables, and they can work as groups."

As the authors continued to examine these quotes, they found that he explained that the large class size was a result of administrators wanting more students in each classroom.

> *[It comes] from the level of the Dean to the Provost. They just want the numbers in the classroom. and that's where it becomes difficult... [The department chair is] saying we need to run up the enrollment you've got to teach in the lecture hall with... That's not the chair's fault, that comes from the Dean and Provost level.*

One of these administrative considerations that Timothy mentioned was financial, running the school like a business, tying back into Bennett's ideas. "And now they just, oh, it's run like a business. You can call me an [old man]. But that's not the way the university should be run." Timothy expressed that administrators wanted to transform schools from teaching-focused to research-focused.

> *In the beginning [my school] was really just teaching focused. But over the years it changed where you know now they promote themselves as the largest research university in the state... They really brought in a focus on research, and then management styles changed from academic to business.*

As the schools transition towards more research, Timothy was concerned that these research teams, too, get taken over by a more managerial style.

> *I worked with a [colleague] who did education research... And the active physics and the problem-based learning, that was great... Unfortunately... they brought in a manager to be the official PI... retired military, and great [at] managing the money. But he didn't help at all.*

According to Timothy, the quality of instruction is not even relevant, as long as someone is in the classroom teaching. "Evaluation doesn't matter, they just need a warm body to [watch] the classes. It's numbers you've got to have, they want enrollment up."

Timothy summarized his perspective on both the original research question ("Why do physics faculty choose lecture?") and the process that leads to faculty being powerless over their pedagogy in an email as follows:

> *I believe your fundamental question is flawed. You say you want to "find out why higher education physics (and other STEM) faculty members prefer the use of lecture." You are assuming that the preference is from the faculty. Once again, the root of the problem is* *money. Standard lecture format is the most cost effective (i.e., cheapest) means available, and university administrators are most interested in the accumulation of tuition dollars. Your choosing to blame faculty is a bias in itself.*

Bennett and Timothy's views have been merged into a single combined theory of what these P&A faculty members feel is the source of their powerlessness to teach using any pedagogy other than lecture.

This theory was expressed as a flowchart, which went through three main revisions, as discussed in Section 2. The original version showed the perceived role of faculty members and administrators, as shown in Figure 1. However, this version ignored the perceived role of the students, so Figure 2 was developed adding their influence on the process, also expanding upon the factors influencing why schools need more money. The third and final version, shown in Figure 3, adds more nuance to the complex interaction that faculty members perceive to exist between the school's need for money, the faculty being one of those expenses, and class size.

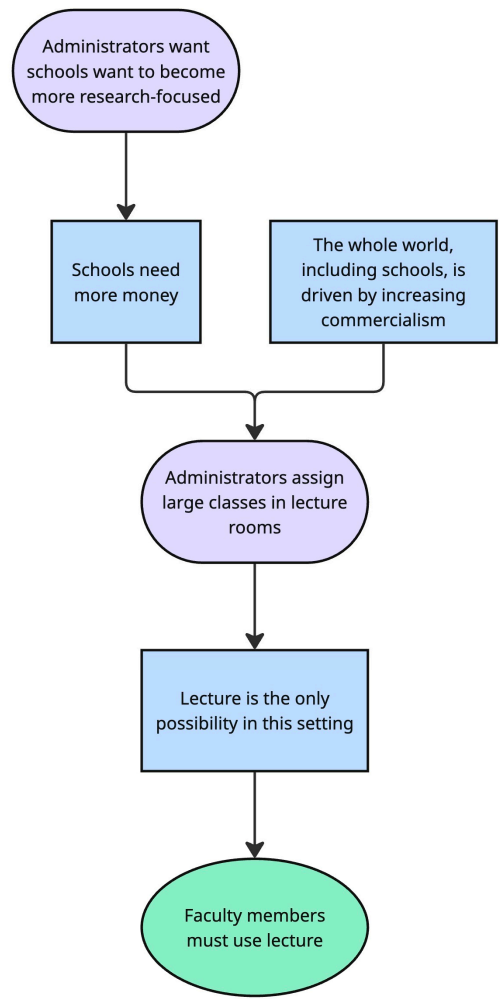

**Figure 1.** The proposed theory of why faculty feel they are forced to use lecture as their pedagogy: the need for money drives administrators to assign large classes, and lecture is the only option; therefore,

faculty members use lecture. Purple rounded rectangles indicate where people are able to make choices in this process (administrators and students); blue rectangles indicate where external forces or consequences without direct human involvement; green oval indicates where people deal with the consequences of others' decisions.

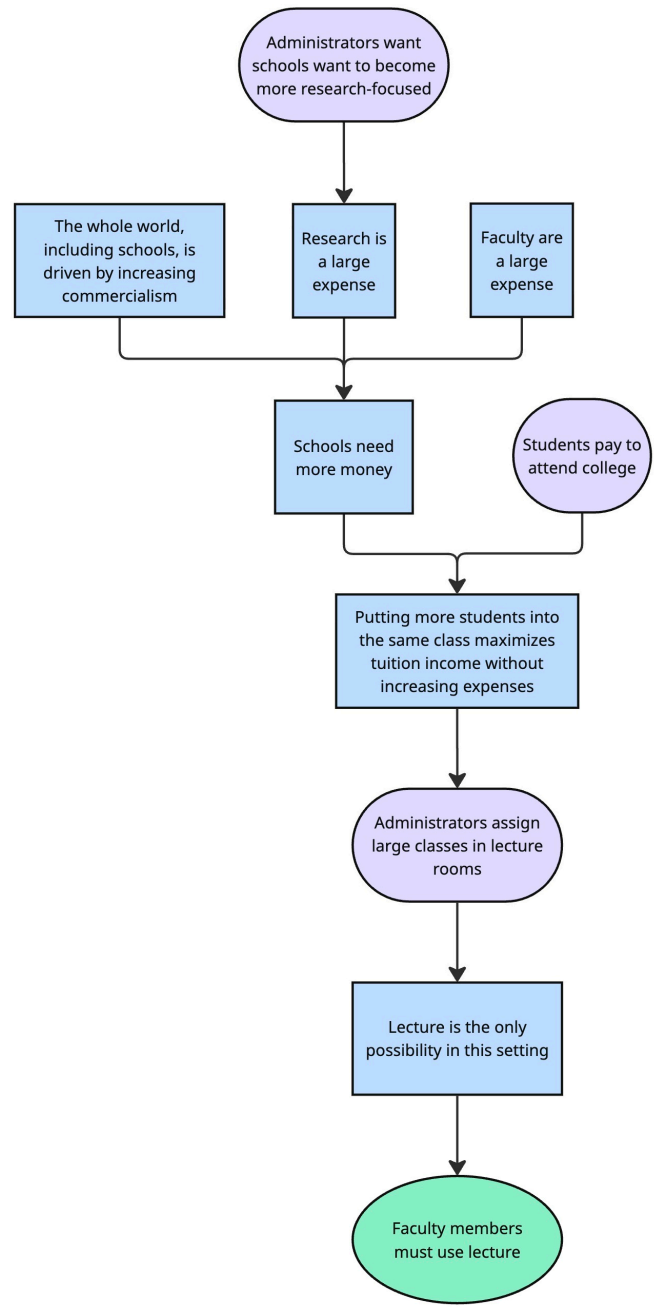

**Figure 2.** In the second graphical version, the role of students is added (paying to attend), as well as the factors influencing the schools' need for money (commercialism, research, faculty). Colors and shapes are the same as in Figure 1.

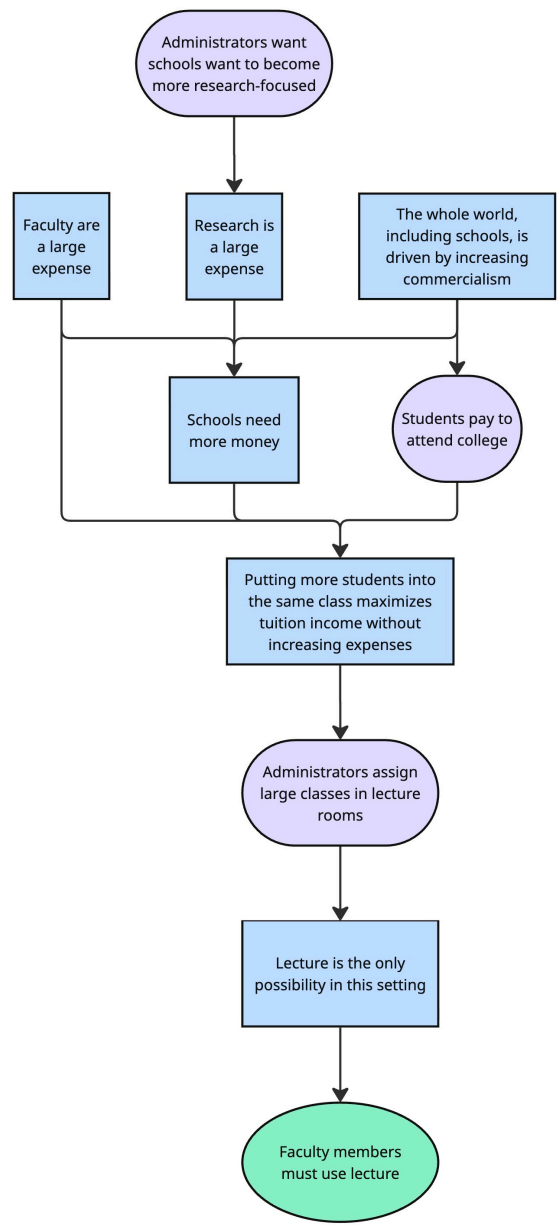

**Figure 3.** The final version of the proposed theory of why faculty feel forced to use lecture pedagogy, which adds arrows indicating the interrelationship between the cost of faculty and that having more students per class reduces financial burdens. Colors as in Figures 1 and 2.

This third version of the theory held by Bennett and Timothy was also expressed as follows:

> *The school needs more money, so administrators choose large class sizes to maximize profit (students) and minimize expense (faculty), large classes inevitably result in large lecture halls, and large lecture halls inevitably result in primarily using lecture. As a result, primarily using lecture in class is not actually a choice for you, it is an inevitable consequence of other people's decisions.*

*3.3. Additional Theoretical Sampling: Email to Participants*

To further refine whether the view of powerlessness is common in P&A faculty members, the researchers continued with the theoretical sampling by sending the text version of the grounded theory to the eight interviewed participants in a follow-up email from the first author. The full text of the email read as follows:

> *One theme I found in a number of the participants was that they felt powerless over much of what happened in their teaching, and that this was driven by financial considerations by the school administrators: the school needs more money, so administrators choose large class sizes to maximize profit (students) and minimize expense (faculty), large classes inevitably result in large lecture halls, and large lecture halls inevitably result in primarily using lecture. As a result, primarily using lecture in class is not actually a choice for you, it is an inevitable consequence of other people's decisions.*
>
> *Would you say that this accurately describes what you have experienced in your teaching?*

Timothy replied that yes, this was an accurate description, with only the tweak that students represent "income" rather than "profit." He then continued to critique the pay for educators at all levels, and more so for non-tenure faculty, "it's all interrelated in my opinion." He agreed that finances are a large driver in how all of education operates, including down to the level of his own classroom.

Bennett's reply disagreed with the teaching connection to finances: "This [infrastructural] aspect is not connected with quality of teaching but budget." Although Bennett denied a connection between budget and "quality of teaching," he did not define what he meant by the "quality of teaching." His reply did not directly address pedagogy. However, his sense of powerlessness over pedagogy was not expressed as it was previously stated in his interview.

Kevin, Lonnie, and Sam all replied that their classes are generally small, and they have not experienced any correlation between class size (or budget) and pedagogy choice. Sam reiterated that he primarily teaches via lecture regardless of class size. Kevin reported that the largest factor is the sizable amount of content he covers as a prerequisite for other courses.

The text description of the powerlessness theory was also presented to the three individuals (Wesley, Cody, and Andromeda) who had not initially expressed any sense of powerlessness to continue the theoretical sampling. Importantly, although not originally articulated, all three individuals reported some level of agreement with the powerlessness concept.

Wesley said that his small class size made it easier to implement student-centered pedagogies such as "Think–Pair–Share" and to use demonstrations. He agreed that, when he is assigned larger rooms for his classes, demonstrations are more difficult to implement.

Cody said his administration required a transition to large class sizes; however, they assessed the impact this had on student learning. His department found that, when student-centered pedagogies were adopted, "such as including smaller recitation sections... there were no measurable learning losses and even some gains; in other cases, they were not effective, and the administration is allowing us to switch back to smaller sections."

Andromeda had previously reported using a mix of student- and teacher-centered pedagogies. In reply to this email, she agreed that she had no control over class size and that it was easier to implement group work in a smaller class size than in a larger class size. However, she did still use some student-centered pedagogies even in larger lectures and felt that increasing class duration was perhaps more impactful than decreasing class size.

*3.4. Additional Theoretical Sampling: Previous Work*

With this new grounded theory in place, the authors returned to data previously gathered during other projects to determine whether the sense of powerlessness permeates other aspects of physics pedagogy. In Schwortz et al. (2017), "another professor" was reported as ascribing to a teacher-centered pedagogy and that he "expressed distrust in the results of educational research" (p. 641). From recollection of the extensive discussions between Schwortz and this faculty member, he expressed that physics faculty have no agency in the process of education research and therefore no agency in what pedagogies are considered "correct." While he did not directly mention what agency he had (or did not have) over his classroom pedagogy, he did clearly discuss concerns that student-centered pedagogy was inappropriate for his classroom and that education researchers did not listen to the concerns of the faculty members in the content areas. This sentiment reemphasizes the need for a study such as this where P&A faculty voice leads the data collection, analysis, and interpretation.

## 4. Discussion

The discovery of a theme of powerlessness over the choice of pedagogy was entirely unexpected, based on the previous literature and completely emergent from the data that the authors collected. A sample of eight post-secondary P&A faculty members, self-selected as primarily using teacher-centered lecture pedagogy, were asked questions related to why they chose to use lecture. Every one of them said that they did not actually choose the teaching style. The exact factors to which they attributed this lack of choice varied. As shown in Section 3, some participants expressed a complex interaction of finances and class size (Bennett and Timothy) or agreed that class size was an important factor (Andromeda, Cody, and Wesley). Others indicated that class size was not the main or only factor but instead the amount of content of the course (Kevin), the resources available at the school (Kevin), or duration of the class period (Andromeda).

From the literature, each of the seven anticipated themes (as described in Section 1) has an approach to solve the challenge, such as using lecture tutorials or TIPERs (Hieggelke et al., 2013). Other approaches include professionally published workbooks designed to make lectures more engaging (to save time for the instructor and avoid the need for training) or addressing the need for student learning rather than entertainment. For this study, as a sense of powerlessness has not been sufficiently studied, approaches to address it are not solidified. Providing ideas for P&A faculty such as reflecting on their reasons for pedagogy use, might be a first step in gathering important P&A faculty voices for a rich data set.

Additionally, rather than convincing faculty to pick something other than a traditional lecture, nudging faculty to understand their place in choosing pedagogy is vital. Advocates of student-centered learning need to show P&A faculty what an active classroom, or mixed active and lecture classroom, can look like (e.g., Malicky et al., 2007), even before convincing them that they want to try something other than pure lecture. Then convincing them, in turn, comes before professional development and other training. In this study, the challenge seems to be persuading the participants that the choice even exists in the first place, but there are steps that could reverse this sense of powerlessness.

Recall that the initial research questions were designed to elicit responses about the seven a priori themes already identified in the literature and not about powerlessness. In response to these questions, Section 3 showed that five out of eight participants organically brought up a lack of agency. But when directly prompted with the theory of pedagogical powerlessness, all the remaining participants agreed with at least one aspect. Thus, 100% of participants in this study had at least some sense of powerlessness over the pedagogy they implemented in the classroom. While the theme of powerlessness was only originally introduced by five out of eight (62.5%) of participants, the fact that 100% agreed with at least one aspect of the theory is crucial. It is possible that the many P&A faculty who continue to use teacher-focused pedagogies do so at least in part due to a feeling that there is no other option available in their specific circumstance.

As a reminder, the purpose of this article is not to examine the reality of this sense of a lack of control over pedagogy but to describe the individual sense of P&A faculty members. However, the faculty members' concerns do agree with what has been found within the context of sociology. The social structures of academia have been found to put constraints upon individual faculty members' choices (referred to as intellectual closure) and that these constraints include the prioritization of teaching versus research (Subramaniam et al., 2014). Furthermore, researchers have expressed concerns about the influence of finances on academia since at least the 1900s, with the power of decisions residing in those who have financial influence (Scott, 2019). However, these sorts of institutional power structures are highly resistant to change (Dacin et al., 2002; Guizzo, 2024). Even if the P&A faculty members are correct and they are powerless to choose pedagogy, the full social context of academia is not something that can be influenced directly through professional development. Changing faculty members' pedagogical choices would involve alternations in academic, social, and other constructs.

This challenge of convincing P&A faculty members to use student-centered pedagogy is further complicated by their feelings that educational researchers are ignoring the multiple aspects of the reality of physics education faculty. The interviewed faculty believe that

1. Physics education already does include teaching styles other than lecture due to including labs and recitation sections with group problem solving;
2. Physics was at the forefront of many educational reforms, such as adding clickers/personal response systems into the classroom, using the flipped classroom pedagogy (introducing new material outside class time, and using class time for practice), and studio physics pedagogy (seamlessly flowing between lecture to introduce new ideas, group problem solving, and mini-lectures or demonstrations that sometimes involve inquiry rather than confirmatory exercises);
3. Educational research does not necessarily address the concerns of physics faculty members, a meta-level of powerlessness.

Physics and astronomy faculty members are proud of their teaching efforts, all without significant training in education research or instructional design and sometimes resent external researchers saying that they are not doing enough. As an example of this, Timothy's interactions with the first author (and interviewer) began antagonistically, as he assumed this study's goal was for educational researchers to find another way to criticize physics faculty members for intentionally not doing enough to help their students. It was only through a lengthy discussion of the research goals, as well as establishing rapport through shared aspects of their backgrounds, that the author-interviewer was able to get him to collaborate and share his thoughts without the initial antagonistic interactions. The "other professor" in the previous work described also expressed that educational researchers do not understand what P&A faculty members are doing or how their classrooms function.

If educational researchers do not take the time to make sincere faculty connections and show that they want to hear the voices of all faculty, even including those who "only" lecture (or appear to go against best teaching practices), other faculty members like Timothy may never speak up at all. Similarly, people providing professional development need to be willing (and have sufficient time available) to reach out to faculty members who do not usually attend PDs and offer to listen to and meet them. Whether physically or in terms of addressing their concerns (such as factors ranging from physical space to number of students, and beyond), connecting with disciplinary faculty in educational research is a critical step to obtaining more accurate data for analysis.

*4.1. Limitations*

This study represents the early stages of developing a grounded theory. Additional work is required in the future to confirm this sense of powerlessness with additional faculty members. It also represents a snapshot of a small number of individuals at a single point in their careers; however, a longitudinal study of faculty members at different stages of their careers may provide different viewpoints.

The five participants who initially expressed powerlessness were all male. Even with the wider group of eight interviewed individuals, only one was female. Six of the eight participants were white. While this reflects the preexisting demographics of P&A faculty in the USA (from which the majority of participants were drawn), it is possible that individuals of different demographics would have different viewpoints. Other participant demographics of the participants were not investigated in this study but could also affect how empowered individuals perceive themselves to be.

There is evidence that the authors' own demographics and history influenced what topics and themes the participants were willing to discuss. For example, one participant noted aloud early in the interview that the interviewer's last name indicated they were of the same ethnicity as the participant, and another participant commented on the interviewer's history as found online. One of the authors is a humanities undergraduate and a white man. Another is an educational faculty member with expertise in science education/partnerships and a white woman. And another is a P&A faculty member with experience in educational research and is nonbinary and multiracial but is often mistaken for a white woman. The authors are situated in a space between educational research situated within an education department and discipline-based education research (DBER, specifically P&A education research, PER and AER) situated within a content area department (specifically, physics and astronomy). Due to this diversity among the authors' backgrounds and crosschecking results, we expect that the authors' demographics and backgrounds are less likely to have affected the analysis process. Lastly, the theoretical sampling was limited. No additional interviews were conducted with any new participants.

*4.2. Future Work*

Grounded theory in action is an iterative process, and, as such, additional investigation on the subject is always needed. New participants could be solicited to determine their perspective on their agency regarding their choice of pedagogy, with interview questions focused on the grounded theory results. It is possible that the ubiquity of this sense of powerlessness (found with all eight interviewed participants) will translate to the vast majority of post-secondary P&A faculty, and they will express a sense of powerlessness over their choice of pedagogy.

Investigations can also be made specifically comparing those individuals who teach via student-centered methods to those who teach via teacher-centered methods to begin to determine how this sense of powerlessness might be mitigated. An action research project might be of particular utility should any faculty who feel powerless wish to address this situation. Future studies could investigate the original seven anticipated factors and themes previously proposed in the literature and additionally add powerlessness.

Although the interviewees in this study were post-secondary faculty members, K-12 primary and secondary educators can face similar challenges when school districts expand class sizes or combine multiple schools, whether due to a lack of funding or due to a lack of teachers. Additional work with secondary P&A teachers may expand the grounded theory suggested herein, or it may show its flaws and alternative approaches. While this study focused on P&A post-secondary educators, it could be enlightening to investigate whether other STEM educators also exhibit similar concerns about a lack of agency.

## 5. Conclusions

At the outset of this project, the researchers did not expect that P&A faculty members would feel a lack of agency in the decision of their classroom pedagogy. The authors include two career education researchers (one of whom is a P&A faculty member, and another a K-12 STEM teacher educator) and a student who wishes to become a teacher in the future. The authors also have extensive experience both in the classroom and working with other STEM instructors. From past experience, the authors anticipated that this study's interviewed faculty members would perceive themselves as the leaders of their classroom and thus the deciding factor of pedagogy. All the literature reviewed discussed factors that influenced instructors' "choice" of pedagogy, without any mention that it might not be a choice after all. Therefore, the authors expected to perform coding for a priori themes suggested in the literature (such as faculty feeling that lecture takes less time to prepare and they needed to spend time on research, that they had never been trained on student-centered pedagogies, or that they choose to lecture because they found it gratifying to have students watching their finely honed performance).

The reality of the data and analysis defied the authors' expectations. More than half of the post-secondary P&A faculty interviewed brought up their sense of powerlessness entirely unprompted, in response to questions on other topics. Two expressed a highly developed theory in their interviews that the original driver of this lack of power was the administration's need to keep finances afloat, which motivated them to increase the student-to-faculty ratio in classrooms, and the faculty then perceived that this large classroom size (both in terms of number of students, and physical seating) required them to teach via the teacher-focused pedagogy of lecture. Each of the eight interviewed faculty members either directly expressed this lack of agency in their interviews or in response to an email describing this theory.

To reiterate: A self-selected sample of P&A faculty who self-identify as using teacher-focused pedagogies universally agreed that they did so because they had no choice. While this was a small sample size, the fact remains that powerlessness is an important factor influencing faculty members to use lecture. The authors wonder if the vast majority of all post-secondary P&A faculty also experience this sense of powerlessness. It is possible that this feeling of powerlessness could be a key factor in faculty members continuing to use lecture rather than more accepted active learning, also known as student-centered pedagogies. Thus, further studies are needed to expand the understanding of faculty members' sense of agency (or powerlessness) over their pedagogy.

The authors would like to provide some final words of guidance for teacher-educators, academic instructional designers, and educators working in college and university centers for teaching and learning. When attempting to reach P&A faculty members, educators should be aware of P&A faculty members' sense of pride and their potential sense of powerlessness over pedagogy. Educational researchers might reach further audiences if they could acknowledge the terms as used by P&A faculty members and incorporate well-known physics pedagogies into studies and lessons. Researchers could also ameliorate their sense of powerlessness by discussing techniques and tools that could be used within large classes of students and lectures in all spaces. Understanding that some educators may feel powerless in their choice of how to teach is an important step toward empowering all educators to freely decide the best teaching methods for their classrooms.

**Author Contributions:** Conceptualization, A.C.S. and A.C.B.B.; methodology, A.C.S. and A.C.B.B.; validation, A.C.S. and M.F.; formal analysis, A.C.S. and M.F.; investigation, A.C.S. and M.F.; resources, A.C.S. and A.C.B.B.; data curation, A.C.S.; writing—original draft, A.C.S.; writing—review and editing, A.C.S., M.F. and A.C.B.B.; visualization, A.C.S.; supervision, A.C.S.; project administration, A.C.S. and A.C.B.B.; funding acquisition, A.C.S. and A.C.B.B. All authors have read and agreed to the published version of the manuscript.

**Funding:** This research study was not grant-funded.

**Institutional Review Board Statement:** This study was conducted in accordance with the Declaration of Helsinki, and the protocol was approved by the Institutional Review Boards at Quinsigamond Community College (Protocol code: none) on 21 August 2024, and College of the Holy Cross (Protocol code: 2025.09.15.Schwortz.v1) on 15 September 2025.

**Informed Consent Statement:** Informed consent (written and/or verbal) was obtained from all subjects involved in this study to participate in and publish this research.

**Data Availability Statement:** The data presented in this study are held by the first author and are not available to the public due to the need to protect the privacy and identity of the participants.

**Acknowledgments:** We want to thank the volunteer study participants. During the preparation of this manuscript/study, the authors used Bing GPT-4o to generate the graphical abstract; the authors have reviewed and edited the output and take full responsibility for the content. We acknowledge support from our institutions in the form of office space, Zoom subscriptions, and computer access. We thank the original stewards of the land we are working on, including the Nipmuc and Massachusett peoples.

**Conflicts of Interest:** The authors declare no conflicts of interest.

## Appendix A

Interview Protocol for Physics Lecture

Interviews are intended to be adaptable.

The questions contained herein are a guidance but not a fixed requirement. Interviewers are encouraged to ask follow-up and exploratory questions to elicit more information from the interviewee. Items in square brackets are intended as guidance for the interviewer, and items not in square brackets are intended to be asked of the participant.

[Hit record to the cloud.]

[Tell a bit about myself, my name, why I'm doing this study, what I hope to learn from this study.]

[Consent form confirmation.]

[Remind the interviewee that they have the right to not answer any question asked, or to terminate the interview at any time. If being video recorded, they also have the right to request the video camera be shut off for any question asked, or for the video camera to be shut off for the remainder of the interview at any time.]

1. [Record the participant's preferred pseudonym and real name; if the participant does not have one, then ask, or move on.]
2. Could you tell me a bit about your current faculty position?

    (a) Full-time/Part-time
    (b) Adjunct/non-tenure/tenure track
    (c) Type of school: community college, teaching-focused, research-focused
    (d) How many classes do you typically teach per semester?
    (e) How many hours do you typically spend in the classroom (including lecture, lab, and discussion) each week?
    (f) What is your typical class size?

  (g) What is the physical space like for your classes? [lecture hall, small classroom, lab space…]
  (h) What courses do you teach? What is the level of these courses? e.g., majors, non-majors/gen ed, seminar style, advanced undergraduate, graduate.
  (i) In what modalities or settings of class do you teach? [E.g., Lecture in a big hall with hundreds of students, small room with 10 students, lab, discussion section, online asynchronous (no meeting times, generally using either pre-recorded videos or written materials), online synchronous (specific meeting times, generally using Zoom or similar).]

3. How long have you been teaching both in total, and at your current institution?
4. How would you describe your teaching style?
   [Look for both a name of a pedagogy, and description of what they're actually doing, e.g.,]

  (a) You said you use pedagogy, can you describe what that term means to you?
  (b) And what does your classroom actually look like? For example, what are the students doing, what are you doing?
  (c) Do you use the same teaching style for all courses you teach? [e.g., non-majors vs. majors, undergraduate vs. graduate]

5. What do you see as the purpose or goal of teaching?

  (a) Is this different depending on what class or what level of student? [e.g., non-majors vs. majors, undergraduate vs. graduate]
  (b) Do you see science education as learning a collection of facts or a process/skill, and why?

6. What do you enjoy about teaching?
7. What do you not enjoy about teaching?
8. [Authoritarianism, "sage on the stage vs. guide on the side", who is in charge of the classroom?]
9. What teaching style do you think your students prefer?

  (a) What makes you conclude they prefer that style? [Evidence]
  (b) Why do you think your students prefer that? [Cause]

10. What teaching style do you think most promotes student learning and why?
11. What teaching style do you think your supervisor prefers?

  (a) What makes you conclude they prefer that style? [Evidence]
  (b) Why do you think they prefer that? [Cause]

12. What do you think are the strengths and weaknesses of teaching via lecture?
13. Do you know of any other teaching styles (pedagogies), and if so, what are their strengths and weaknesses? [Examples: studio physics/MIT TEAL, flipped classroom, inquiry-based learning, project-based learning (PBL), case studies, etc.]
14. How are you assessed on your teaching?
    [Listen for the following possibilities, and if they don't mention them, then bring it up:]

  (a) Student evaluations
  (b) Peer observations
  (c) Supervisor observations
  (d) Teaching and learning center observations as evaluation
  (e) Student grades or pass rate
  (f) Self evaluation

15. What other commitments do you have besides teaching?
    [Follow-ups:]

   (a)   What is the priority order of these commitments?
   (b)   Why is that your order?

16. What teaching style or styles were used in your own education?
    [Follow-ups:]

    (a)   Undergraduate, aka post-secondary or tertiary school?
    (b)   Graduate? (Master's or PhD coursework)

17. What types of training in how to teach (pedagogy instruction) have you received?
    [Examples:]

    (a)   Grad school teacher training workshop, 1 day to 1 week
    (b)   Departmental grad school workshop, 1 day to 1 week
    (c)   Departmental grad school mentoring
    (d)   Course on teaching
    (e)   Thrown into lab and didactic

18. I have a few questions about demographics, if you don't mind answering. If you would rather not, that's fine.

    (a)   In what country did you complete most of your education?
    (b)   What is your gender?
    (c)   What is your race, ethnicity, or cultural background?
    (d)   What is your age?
    (e)   Do you have children?

   "Thank you so much for your time, that is all the questions I have for now. Do you have any other questions or comments for me?"